\documentclass[11pt,twoside]{article}
\usepackage{./asp2014}
\newcommand{\teff}{$T_{\rm eff}$}
\aspSuppressVolSlug
\resetcounters

\begin{document}

\title{Spectral analysis of the sdO standard star Feige 34}
\author{Marilyn Latour,$^1$ Pierre Chayer,$^2$ Elizabeth M. Green,$^3$ and Gilles Fontaine,$^4$
\affil{$^1$Dr. Karl Remeis-Observatory \& ECAP,
               Friedrich-Alexander University Erlangen-N\"{u}rnberg, Bamberg, Germany; \email{marilyn.latour@fau.de}}
\affil{$^2$Space Telescope Science Institute, Baltimore, Maryland, USA; \email{chayer@stsci.edu}}
\affil{$^3$Steward Observatory, University of Arizona, Tucson, Arizona, USA; \email{egreen@email.arizona.edu}}}
\affil{$^4$Universit\'{e} de Montr\'{e}al, Montr\'{e}al, Qu\'{e}bec, Canada; \email{fontaine@astro.umontreal.ca}}

\paperauthor{M. Latour}{marilyn.latour@fau.de}{0000-0002-7547-6180}{Friedrich-Alexander University Erlangen-Nuremberg}{Dr. Karl Remeis-Observatory \& ECAP, Astronomical Institute}{Bamberg}{}{96049}{Germany}
\paperauthor{P. Chayer}{chayer@stsci.edu}{}{Space Telescope Science Institute}{}{Baltimore}{Maryland}{}{USA}
\paperauthor{E. M. Green}{egreen@email.arizona.edu}{}{University of Arizona}{Steward Observatory}{Tucson}{Arizona}{}{USA}
\paperauthor{G. Fontaine}{fontaine@astro.umontreal.ca}{}{Universit\'{e} de Montr\'{e}al}{}{Montr\'{e}al}{Qu\'{e}bec}{}{Canada}

\begin{abstract}
We present our current work on the spectral analysis of the hot sdO
star Feige 34. We combine high S/N optical spectra and fully-blanketed 
non-LTE model atmospheres to derive its fundamental
parameters (\teff, log $g$) and helium abundance. Our best fits indicate
\teff =63 000 K, log $g$=6.0 and log $N$(He)/$N$(H)=$-$1.8.
We also use available ultraviolet spectra (IUE and FUSE) to measure
metal abundances. We find the star to be enriched in iron and nickel
by a factor of ten with respect to the solar values, while lighter
elements have subsolar abundances. The FUSE spectrum suggests
that the spectral lines could be broadened by rotation.
\end{abstract}

\section{Astrophysical context}
Despite its brightness and its status as a calibration star,
information on the physical parameters of Feige 34 is rather scarce.
\citet{the91} adopted \teff\ $\sim$80 000 K and log $g$ = 5.0
by comparing its optical spectra to the one of BD+28$^{\circ}$4211. The main goal
of their study was to characterize the infrared excess of the star,
which they suggested is caused by a dwarf companion.
Feige 34 parameters were later updated by \citet{haas97}
to 60~000 K and log $g$=5.2. His results were based on the IUE
spectrum of the star. He also quantified the overabundances of iron
and nickel, to be 10x and 70x solar, respectively, and noted the
similarity between the IUE spectra of Feige 34, Feige 67 and
LS II +18$^{\circ}$9. Since not much is known about the abundance pattern in hot
hydrogen-rich sdOs we felt it is important to revisit such stars and constrain the 
chemical composition of their atmosphere. We first investigated the case of Feige 34 and
report our progress below.

\section{Optical spectra}
For the purpose of this work, we invested a significant amount of time
in the computation of new grids of TLUSTY non-LTE line-blanketed model
atmospheres. They cover the temperature range between 50 and 70 kK
and include solar and supersolar(10x) metallicities. Our models include eight metallic elements that are treated in non-LTE (C,N,O,Mg,Si,S,Fe, and Ni).
Similar models have
been used previously to successfully fit and reproduce the optical Balmer and
helium lines of the hot (\teff\ =82 kK) sdO star BD+28$^\circ$4211 \citep{lat15}.

We fitted three different medium/low resolution
spectra with our new model grids, and obtained the following
parameters: \teff\ $\sim$63~000 K, log $g$ $\sim$6.0 and log $N$(He)/$N$(H) =$-$1.8, when
using the metal-enhanced grid. The best fit for one of the spectra is shown in Fig. \ref{fig1}.
While the quality of the fit is better with the metal-rich grid than with the solar metallicity one
some discrepancies still remain, mostly in the core of H$\alpha$ and H$\beta$.
As for the atmospheric parameters, both grids lead to similar parameters, with \teff\ being slightly lower
(by $\sim$2500 K) when the solar metallicity grid is used. 

\articlefigure[angle=270, width=0.8\textwidth]{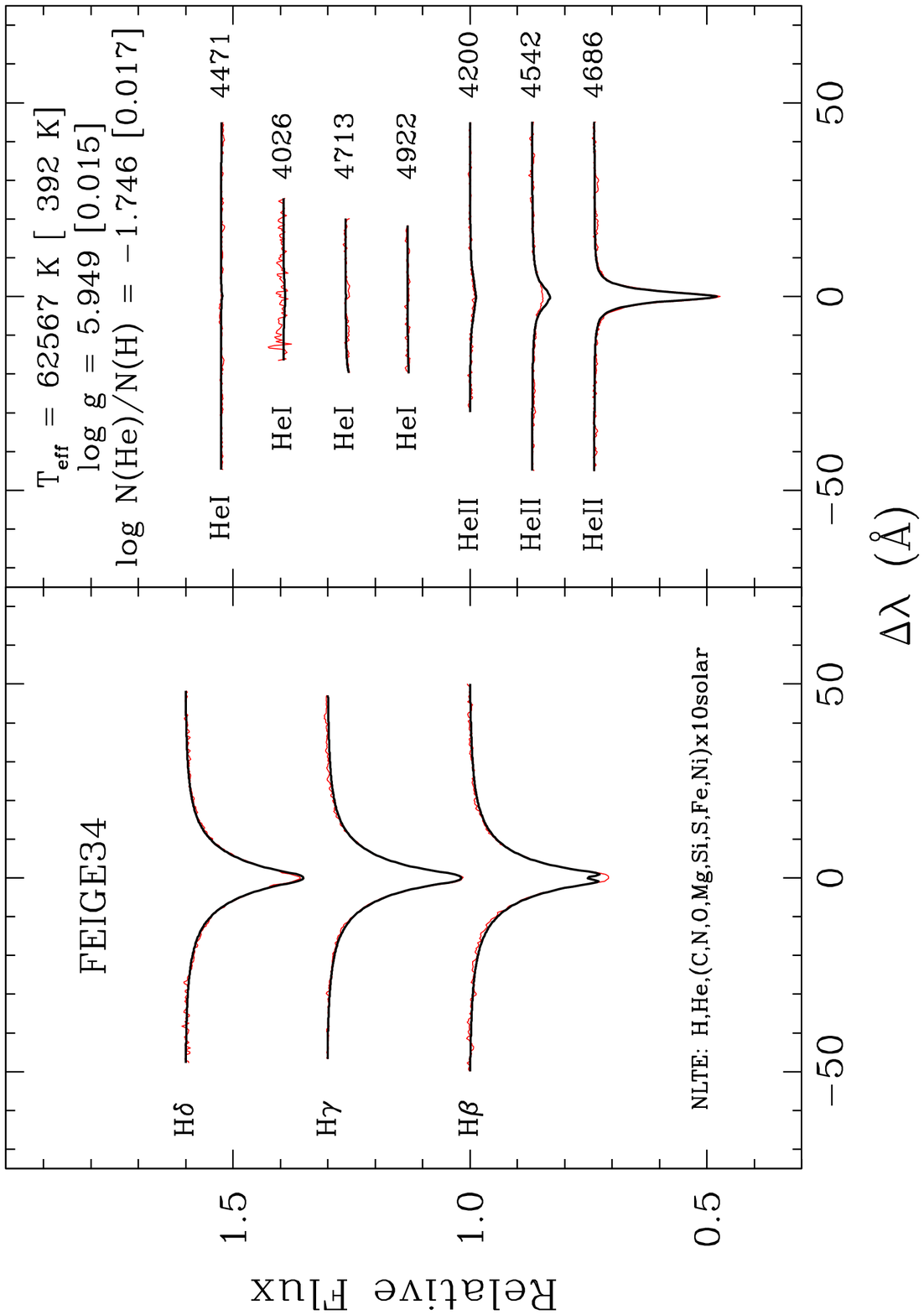}{fig1}{Best fit (black) of the MMT 1\AA\ resolution spectrum (red) of Feige 34}

We inspected the H$\alpha$ and H$\beta$ lines in high resolution HIRES (Keck) spectra. The emission core of these lines is resolved in the HIRES spectra allowing us to compare the observed line profiles with our models. 
There is a small improvement in the reproduction of the line profile by the metal-rich models but both type of models
overestimate the strength of the emission core of these lines, explaining the discrepancies observed in our best fits. This effect is seen in Fig. \ref{fig1} for H$\beta$.

The spectral energy distribution of Feige 34 was also examined, and a
fit of the different magnitudes of the star (from the IUE flux
to WISE) suggests a M-type companion (J. Schaffenroth, priv. comm).
It is however unclear if the system is a physical binary, as no radial velocity variations have been
detected so far in Feige 34.

\articlefigure[angle=90, width=1\textwidth]{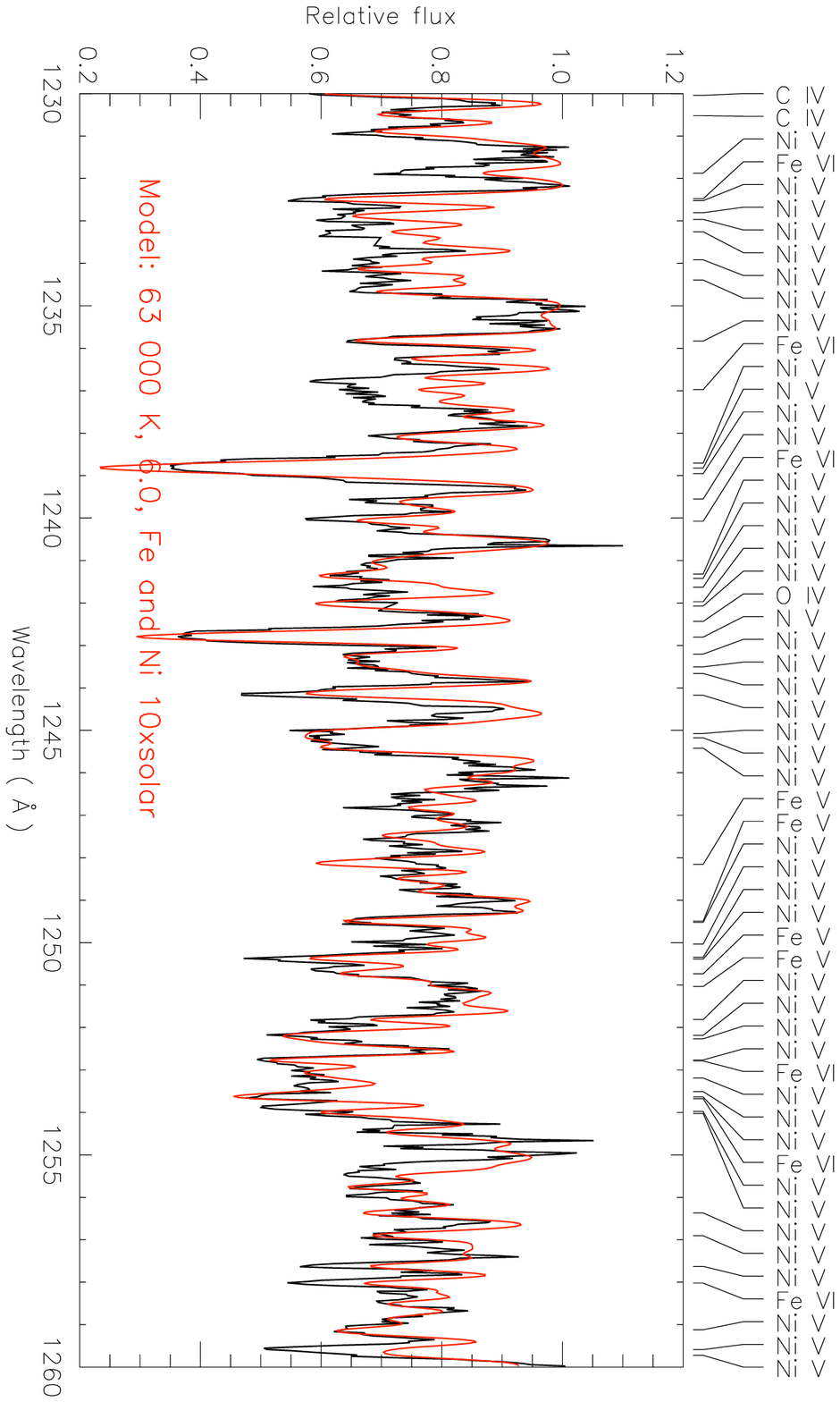}{fig2}{Portion of the IUE spectrum dominated by Fe and Ni lines, also seen is the N~\textsc{v} doublet (1238-1242 \AA). Overplotted in red is a model including Fe and Ni (10x solar).}
\articlefigure[angle=90, width=1\textwidth]{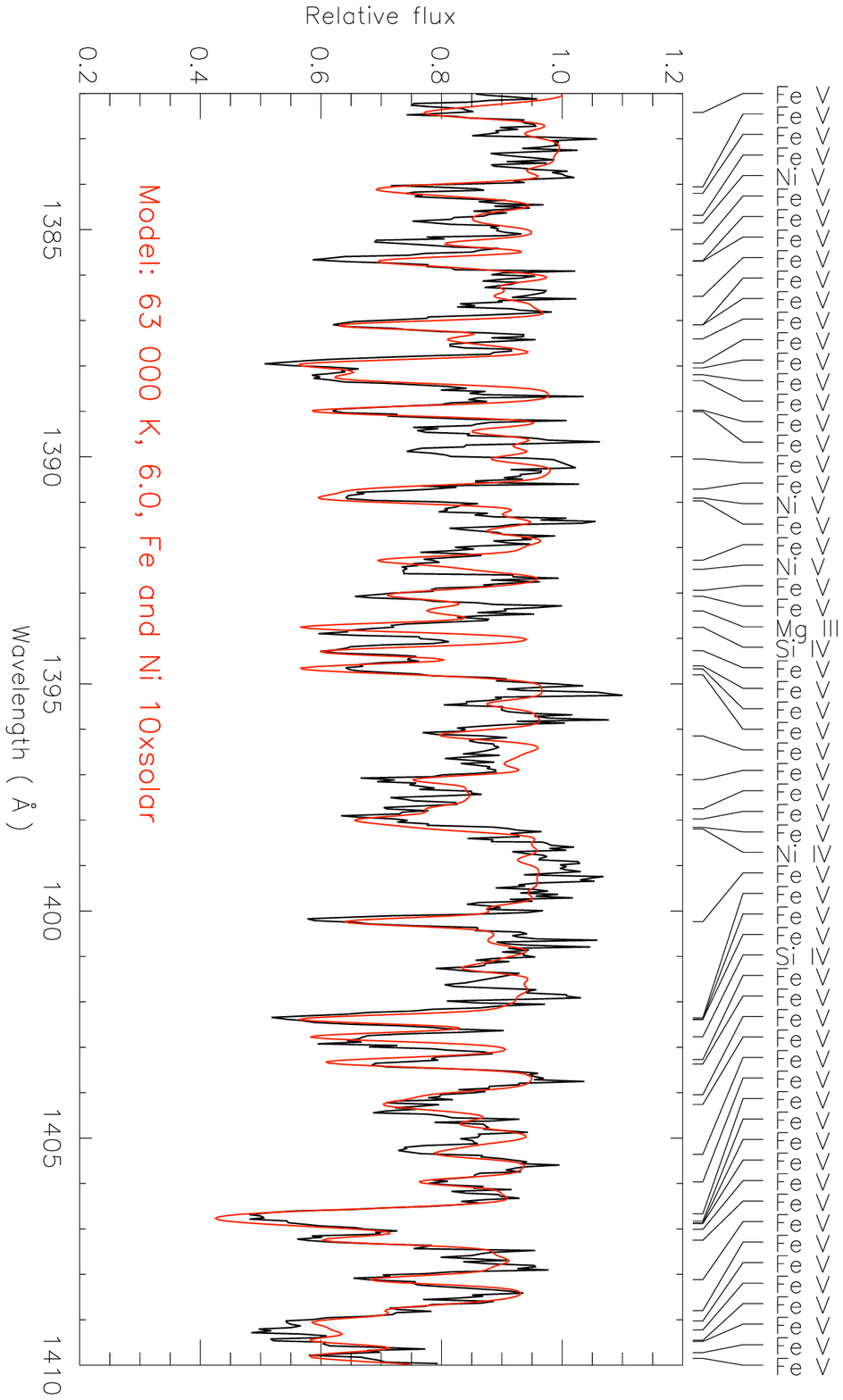}{fig3}{Same as Fig. \ref{fig2} but for a different range rich in Fe~\textsc{v} lines.}

\section{UV Spectra}

Feige 34 has been observed with the IUE satellite as a calibration standard. We retrieved the two high dispersion, short-wavelength spectra available from the archive and combined them. Two sections of the resulting spectrum are shown in Fig. \ref{fig2} and \ref{fig3}. The spectrum is heavily crowded by iron and nickel lines that are mostly blended, due to their strength and the limited resolution ($\Delta\lambda$ $\sim$0.15\AA). Nevertheless, a good agreement between our models and the observed spectrum is achieved in the two regions shown, assuming an enrichment of ten times solar for iron and nickel. Nitrogen, oxygen and carbon lines are also visible in IUE and the abundances derived are reported in Table 1.

\begin{table}[!ht]
\begin{center}
\caption{Measured abundances in Feige 34}
\smallskip
{\small
\begin{tabular}{lcc}  
\tableline
\noalign{\smallskip}
Element & log $N$(X)/$N$(H)  & log $N$(X)/$N$(H) -  \\
 & ($\pm$0.4 dex) & log $N$(X)/$N$(H)$_{\rm{sol}}$ \\
\noalign{\smallskip}
\tableline
\noalign{\smallskip}
He & $-$1.8 & $-$0.8 \\
C &  $\le$ $-$5.8 &  $\le$ $-$2.2\\
N & $-$5.0 & $-$0.8 \\
O & $-$5.8 & $-$2.5 \\
S & $-$5.5 & $-$1.0 \\
Fe & $-$3.5 & $+$1.0 \\
Ni & $-$4.5 & $+$1.0 \\
\noalign{\smallskip}
\tableline 
\end{tabular}
}
\end{center}
\end{table}

Feige 34 was also observed by the FUSE satellite in a complementary $\lambda$ range (905$-$1187 \AA) to IUE and with a higher resolution ($\Delta\lambda$ $\sim$0.06\AA). However the FUSE spectrum of the star appears to be quite polluted by interstellar lines (mostly H$_2$), rendering photospheric lines identification challenging. 
Nevertheless, the S~\textsc{vi} resonance lines are clearly visible as well as four N~\textsc{iv} lines around 922$-$924 \AA. These N lines are less sharp than expected from synthetic spectra and suggest a rotational broadening of at least vsin$i$ = 25 km s$^{-1}$. Such high rotation is not typical in hot subdwarfs (unless they are in a close binary system) and this aspect will need a more thorough investigation. Unfortunately, the optical spectrum of the star is devoid of metal lines (even at high resolution) and the IUE spectrum is not optimal for such a measure.


\acknowledgements This work was supported by a fellowship for postdoctoral researchers from the Alexander von Humboldt Foundation. Computation was supported in part by the NSERC Canada.


\begin{thebibliography}{}
\bibitem[Haas (1997)]{haas97} Haas, S.\ 1997, PhD Thesis,  Friedrich-Alexander-Universit\"{a}t Erlangen-
N\"{u}rnberg
\bibitem[Latour et al.(2015)]{lat15} Latour, M., Fontaine, G., Green, E.~M., \& Brassard, P.\ 2015, \aap, 579, A39 
\bibitem[Thejll et al.(1991)]{the91} Thejll, P., MacDonald, J., \& Saffer, R.\ 1991, \aap, 248, 448 


\end{thebibliography}


\end{document}